\documentclass[11pt]{article}

\RequirePackage{amsthm,amsmath,amsfonts,amssymb}
\RequirePackage[authoryear]{natbib}
\RequirePackage[colorlinks,citecolor=blue,urlcolor=blue,backref=page,backref=page]{hyperref}
\RequirePackage{graphicx}

\usepackage{siunitx}
\usepackage{booktabs}
\usepackage{longtable}
\usepackage{multirow}
\usepackage{rotating}

\theoremstyle{plain}

\newtheorem{theorem}{Theorem}[section]
\newtheorem{proposition}{Proposition}[section]
\newtheorem{corollary}{Corollary}[section]
\newtheorem{lemma}[theorem]{Lemma}
\theoremstyle{definition}

\theoremstyle{remark}

\title{Here Be Dragons: Bimodal posteriors arise from numerical integration error in longitudinal models}
\author{Tess O'Brien, Matthew T. Moores, David Warton, and Daniel Falster}

\begin{document}
	\maketitle
		
	\begin{abstract}
		Longitudinal models with dynamics governed by differential equations may require numerical integration alongside parameter estimation. We have identified a situation where the numerical integration introduces error in such a way that it becomes a novel source of non-uniqueness in estimation. We obtain two very different sets of parameters, one of which is a good estimate of the true values and the other a very poor one. The two estimates have forward numerical projections statistically indistinguishable from each other because of numerical error. In such cases, the posterior distribution for parameters is bimodal, with a dominant mode closer to the true parameter value, and a second cluster around the errant value. We demonstrate that multi-modality exists both theoretically and empirically for a linear first order differential equation, that a simulation workflow can test for evidence of the issue more generally, and that Markov Chain Monte Carlo sampling with a suitable solution can avoid bimodality. The issue of multi-modal posteriors arising from numerical error has consequences for Bayesian inverse methods that rely on numerical integration more broadly.
	\end{abstract}

	\section{Introduction}
	The use of numerical integration in methods for estimating differential equation parameters from observations introduces a source of error \citep{butcher2016numerical} that is distinct from other processes such as measurement error \citep{agapiou2014analysis}. In the Bayesian literature, the effect of numerical error on the posterior distribution is under-explored and we wish to present a case study on identifiability problems which doubles as a procedure for investigating them under similar circumstances. In this paper, we describe a specific pathology arising from numerical error: erroneous multi-modality of posterior parameter distributions for linear first order ODEs where there are multiple parameter combinations that produce the same numerical solution. Such multi-modality is a previously undescribed form of non-identifiability in Bayesian methods.
	
	We first observed bimodality arising from numerical methods while testing a longitudinal model for the von Bertalanffy function \citep{von1938quantitative}. Posterior bimodality leads to independent Markov Chain Monte Carlo (MCMC) chains converging to very different, but very specific, parameter combinations, one which is a good estimate of the true values (\emph{i.e.}\ parameters known from simulation), and a separate very bad estimate. For the erroneous combination, \textit{under the chosen numerical method and step size}, forward projection produces estimates of $Y(t_j)$ that are the same as those produced by the correct parameters, due to numerical error. The posterior is, after all, conditioned on the numerical method used in the model, not the true solution to the integral. In general, where numerical methods are implemented in estimation they should be treated as part of the statistical model. As our case study arose from ecological literature, we are thinking about growth functions for DEs, and growth parameters as what is being estimated. However the example is a simple DE likely to appear in other examples. The language we use will reflect the ecological background but the situation is more general.
	
	Non-identifiability is a general statistical concept for a situation where multiple estimate values have an indistinguishable relationship to data \citep{rothenberg1971identification}. The form of non-identifiability we are dealing with is different to cases described in the literature. For example \cite{gelfand1999identifiability} discusses situations where the set of observations is small relative to the parameter space such as can arise in generalised linear mixed effects models. \cite{auger2016state} addresses large measurement error as a potential source, and \cite{cockayne2019bayesian} discusses a situation where the posterior distribution is bimodal because there are two true solutions to the underlying function. Numerical integration error as a source of estimation problems is rarely addressed in the existing literature. The only mentions we found were \cite{agapiou2014analysis} looking at convergence for MCMC at different levels of discretisation and \cite{cotter2010approximation} where the step size of an Euler method was found to bias estimates. Supplementary material to \cite{obrien2024allindividuals} investigated bias arising from numerical error for a linear ODE. Posterior multi-modality arising from numerical error appears to be completely novel. 
	
	When multi-modality arises from numerical error, stability of the integration method matters for potentially a very large region of the parameter space -- far beyond true or even plausible parameter combinations -- the boundaries of which are probably unknown when first going in. The numerical methods we use (Runge-Kutta 4 and 4-5 order) are extremely common (\emph{e.g.}\ \citealp{falster2017multitrait}), and numerically more reliable than other commonly used methods such as Euler (implemented in \citealp{cotter2010approximation} and \citealp{iida2014linking}). We demonstrate that the problem arises theoretically, and is empirically observable, for a linear first-order ordinary differential equation with negative slope, such as the von Bertalanffy \citep{von1938quantitative} model where we first observed it. The von Bertalanffy model is our central example in a re-parameterised form. Linear ODEs are common, and the RK4 method is typically considered reliable for short step sizes \citep{butcher2016numerical}. To find a situation where the combination of the two produced a novel pathology was unexpected. The linear ODE has an analytic solution which allows me to side-step the identifiability problem by not using numerical methods, however if numerical error introduces non-uniqueness elsewhere it may be in situations without such a work-around. 
	
	In this paper we first prove that in general, linear first order ODEs subject to RK4 numerical integration in a longitudinal model will have multi-modal posteriors. We then present a simulation workflow for matching empirical results that allows a user to test the parameter space of more general models for multi-modal pathologies before embarking on large-scale analysis. Our testing uses MCMC estimation with different true parameter values. Supplements A and B show that different prior parameters and estimation methods do not change the problem. We report the results of experiments to demonstrate how the problem appears in practice, and investigate whether common approaches to dealing with such issues work. We have another function of interest -- the extremely non-linear Canham function based on a log-normal distribution function from \cite{canham2004neighborhood} -- and in Supplement C: Canham Testing we use the simulation workflow to test whether there is evidence of bimodal posteriors for parameters in that case as well. 
	
	Our use of simulation is distinct from a simulation-based calibration method \citep{talts2020validating, modrak2023simulation} which uses data simulated from the chosen prior to test the consistency of the posterior with rank statistics. We are doing a predominantly qualitative rather than quantitative analysis testing for evidence of the existence of posterior bimodality arising from numerical error, and use simulated data that does not come from the chosen priors. We consider this a reasonable approach because our intent is to demonstrate the existence and mechanism by which numerical bimodality arises, we already have a fix in the linear ODE's analytic solution. In the spirit of \cite{gelman2020bayesian}, deliberately breaking a model can be informative, and we aim to present a generally applicable workflow using simulation for testing whether an intended model is susceptible to these novel problems.
	
	\section{Theoretical analysis}
	We demonstrate the bimodal posterior estimate behaviour for dynamics based on the linear first-order ordinary differential equation (ODE)
	\begin{equation}\label{eqn_linearDE}
		\frac{dY}{dt} = f(Y(t), \beta_0, \beta_1) = \beta_0 - \beta_1 Y(t),
	\end{equation}
	under the longitudinal model
	\begin{equation}\label{eqn_hereDragons_Longitudinal}
		Y(t_{j+1}) = Y(t_j) + \int_{t_j}^{t_{j+1}} f(Y(t), \beta_0, \beta_1)\,dt.
	\end{equation}
	
\begin{figure}
		\centering
		\includegraphics[width=0.92\linewidth]{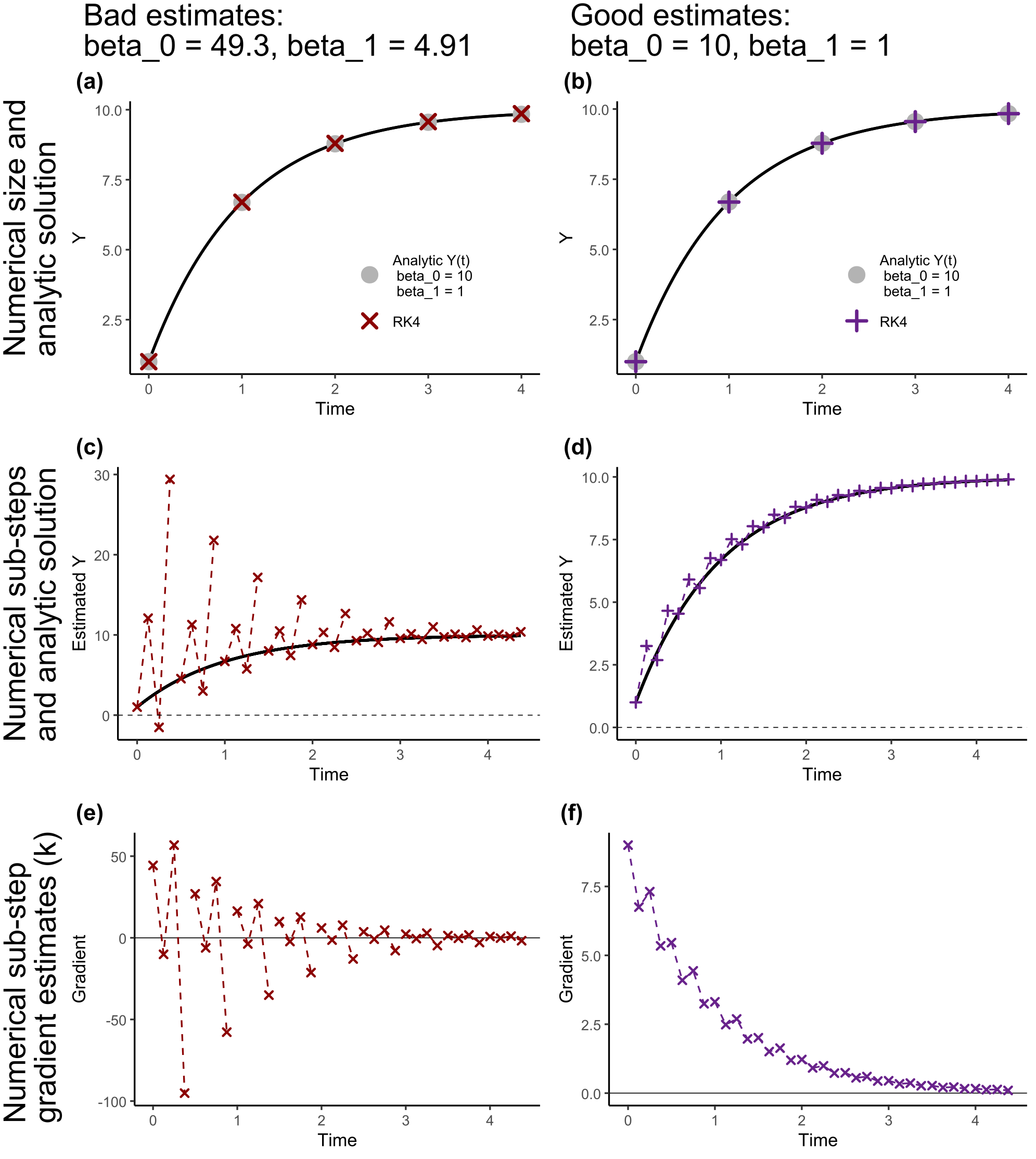}
		\caption{Behaviour of numerical solutions at erroneous and true parameter locations showing considerable instability at the erroneous point. The true $Y(t_j)$ values are those of the analytic solution with $\beta_0=10$ and $\beta_1=1$ that the model attempts to estimate, the erroneous values $\beta_0 = 49.3$ and $\beta_1 = 4.91$ can be found both in theoretical results and through empirical testing. \textbf{(a)} and \textbf{(b)} compare the estimated sizes for integer times between 0 and 4 to the analytic solution to show both the good and bad parameters give numerical values very close to the analytic ones. In \textbf{(c)} and \textbf{(d)} we see the sub-steps as well, which are connected up based on the numerical step they come from. The bad parameter estimates demonstrate considerable instability and even a negative value. \textbf{(e)} and \textbf{(f)} give the gradients at the numerical sub-steps and show why the estimated sizes are so unstable for the bad estimate as the gradient oscillates about 0. }
		\label{fig_Numerics}
	\end{figure}

	Equation \eqref{eqn_linearDE} is considered a particularly pathological example for numerical methods \citep{butcher2016numerical} as it is unbounded above and below, and the negative slope in particular means that negative growth increments can arise from the numerics even if the underlying system is strictly positive. The move from continuous time in the analytic solution to discrete time in the numerical one can result in estimates of $Y(t)$ oscillating around the asymptotic size. The oscillation arises when the calculated gradient overshoots 0 then is pulled back as seen in Figure \ref{fig_Numerics}. Equation \eqref{eqn_linearDE} serves as an ideal example of the bimodal posterior because of the known pathologies, it is simple with two parameters, and has an analytic solution which allows us to easily access true values for $Y(t)$ without introducing compounding numerical error. The known analytic solution for a linear model is
	\begin{equation}\label{eqn_linearDESoln}
		Y(t) = \frac{\beta_0}{\beta_1} + \bigg(Y(0) - \frac{\beta_0}{\beta_1}\bigg) \exp(-\beta_1 t),
	\end{equation}
	which asymptotes to 
	\[\lim_{t \to \infty} Y(t) = \beta_0/\beta_1.\]
	We choose the initial condition $0 < Y(0) = 1 < \frac{\beta_0}{\beta_1}$. Equation \eqref{eqn_linearDESoln} gives a mathematical model that is defined for regions that may not correspond to realistic behaviour in the system being modelled. For example, if $Y$ represents physical size, $Y(t) < 0$ is impossible. Negative growth is also possible if $Y(t)>\frac{\beta_0}{\beta_1}$.
	
	In this demonstration we use a Runge-Kutta 4th order numerical method \citep{butcher2016numerical} with different choices for the step size to show that even `small' step sizes (small enough to have good behaviour at the true parameters) can give problems. A step size in this context is the length of a single iteration of the numerical integral, and is at most the time between observations: $t_{j+1}-t_j$. As we are using synthetic data, we have observations 1 unit of time apart, and take step sizes to be 0.5, 0.25, and 0.125 which divide 1 evenly.
	
	We assume data consists of finite observations of the form $y_{j}$ at time $t_j$ that look like
	\begin{equation}\label{eqn_obs}
		y_j = Y(t_j) + \text{error},
	\end{equation}
	and have some finite level of precision. In the Bayesian literature, this set-up is an issue of practical identifiability, where a finite number of observations combined with measurement error complicate access to the true parameters \citep{latz2023bayesian}. We use finite true $Y(t_j)$ values for theoretical results, and observations with error $y_j$ for empirical ones.
	
	For this demonstration we use added error of $\mathcal{N}(0, 0.1)$ and rounding to produce simulated data with measurement precision of 0.1, analogous to \SI{1}{\milli\metre} measurement precision and standard deviation on observations in centimetres. The scale of measurement is chosen because the method in \cite{obrien2024allindividuals} was developed for use in ecological size data where such structure shows up \citep{condit1998tropical}. We want to demonstrate a realistic measurement process, but the bad numerical behaviour occurs even with a high level of precision, and much smaller error as the underlying cause is mathematical. 
	
	We are attempting to estimate the parameters $\beta_0$ and $\beta_1$ from simulated observations $y_j$ of the form in Equation \eqref{eqn_obs}. We estimate $\hat{Y}_j$ given the prior distribution
	\[y_j \sim \mathcal{N}(\hat{Y}_j, 0.1)
	\]
	and the longitudinal model in Equation \eqref{eqn_hereDragons_Longitudinal}, which encodes the auto-correlation in the data. For this demonstration we are looking at a single simulated individual. To make fitting the model easier we apply a translation by the mean observed size $\bar{y}$, which produces the implemented DE
	\[f(Y(t),\beta_c, \beta_1, \bar{y}) = \beta_c - \beta_1(Y(t) - \bar{y}).
	\]
	The translation does not mathematically change the behaviour of $Y(t)$. We extract estimates for $\beta_0$ by back-transforming $\beta_c$ as by definition
	\[ \beta_0 = \beta_c +\beta_1\bar{y},
	\]
	so the output estimates are consistent with Equation \eqref{eqn_linearDE}. We have default prior distributions for the parameters which are
	\[0 < \beta_1 \sim \log\mathcal{N}(0,2),\qquad \beta_c \sim \mathcal{N}(1,2),
	\]
	where the mean and standard deviation for $\beta_1$ are for the underlying log-transformed normal distribution. We chose a log-normal distribution as it enforces positive values to maintain our parameterisation of Equation \eqref{eqn_linearDE}. A prior sensitivity analysis for both centre and spread is provided in Supplement A and demonstrates that the bimodality is persistent under reasonable prior parameter choice. The minimal hierarchical set-up is based on a single individual structure such as used in \cite{obrien2024allindividuals}, but with a number chosen for the error standard deviation instead of fitting an error parameter.
	
	\subsection{Numerical integration theory}
	In this section we prove non-identifiability of the linear model parameters under a Runge-Kutta 4th order numerical method (RK4) in a theoretical context using the true values rather than observed data. We prove that, amongst the set of solutions for which the asymptotic size is $\alpha$, the linear ODE has multiple $\hat{\beta}_1$ parameters that will produce the same sizes over time for the RK4 method, then look at the specific case study where $\beta_0=10$ and $\beta_1=1$.
	
	For those unfamiliar, the RK4 method uses a weighted average of gradients across an interval to predict the next size \citep{butcher2016numerical}. Given an initial size $Y(t_j)$, fixed parameter values $\beta_0$ and $\beta_1$, and a numerical step size $h = t_{j+1}-t_j$, we estimate
	\begin{equation}\label{eqn_RK4}
		\hat{Y}(t_j + h) = Y(t_j) + \frac{h}{6}(k_1 + 2k_2 + 2k_3 +k4),
	\end{equation}
	where the $k$ values are estimates of the gradient based on the DE
	\[f(Y(t), t, \beta_0, \beta_1) = \frac{dY}{dt},
	\]
	with formulas calculated iteratively
	\begin{align*}
		k_1 =&\ f(Y(t_j), t_j, \beta_0, \beta_1),\\
		k_2 =&\ f(Y(t_j) + k_1h/2, t_j + h/2, \beta_0, \beta_1),\\
		k_3 =&\ f(Y(t_j) + k_2h/2, t_j + h/2, \beta_0, \beta_1),\\
		k_4 =&\ f(Y(t_j) + hk_3, t_j + h, \beta_0, \beta_1).
	\end{align*}
	In the case of Equation \eqref{eqn_linearDE} the time parameter $t$ does not appear explicitly so our $k$ formulas do not depend on it. We will refer to where the $k$ parameters are estimated as `sub-steps'. Note that $t_j$ and $t_{j+1} = t_j + h$ are typically different to the times at which observations occur. The observation interval is an upper bound on the numerical step size $h$, but we can take multiple numerical steps between the observations.
	
	In a chosen situation where $\beta_0=10$ and $\beta_1=1$ which produces a bimodal posterior, at the second (erroneous) parameter combination the numerical method is sufficiently unstable that forward projection produces $\hat{Y}(t_j)$ values that match the analytic solution at the true parameters. For a visual comparison see Figure \ref{fig_Numerics}(a) and (b), which leverages the bad estimates ($\beta_0 = 49.3, \beta_1 = 4.91$) paired to our chosen parameters $\beta_0 =10$ and $\beta_1=1$. The estimation algorithm will converge to \emph{any} parameter combination that gives $\hat{Y}(t_j)$ close to the observations provided. In these cases the algorithm finds a point far from the true value where size estimates match the observations extremely well because of numerical integration error. Non-identifiability here is not a problem with MCMC or another estimation method, it is a structural issue arising from the use of numerical integration in the longitudinal structure.
	
	We can see the problem in plots of the estimated values for $Y(t)$ and $f(Y(t))$ at each step and sub-step for time from 0 to 4. Figure \ref{fig_Numerics} shows instability of the numerical method at $\beta_0 =49.3$, $\beta_1=4.91$ (the bad parameter combination for this example we found both empirically and theoretically), and stability at $\beta_0 =10$, $\beta_1=1$. Both parameter combinations converge to approximately the same asymptotic size $\beta_0/\beta_1 = 10$, so in a sense the long-term numerics are stable, but because we are using the numerical method to fit longitudinal data, we care about the stability at the intermediate times as well. The main feature of interest is the huge variation of the $\hat{Y}$ estimate in sub-steps given the erroneous parameter values, which comes from the gradient oscillating between positive and negative values in Panels (a) and (c). The excessive size happens when the increment estimate at a step $\hat{Y}(t_1)$ puts the next value $\hat{Y}(t_2)$ past the asymptotic limit $\beta_0/\beta_1$, where the gradient is negative. Because the gradient at $\hat{Y}(t_2)$ is negative, the subsequent $\hat{Y}(t_3)$ value will be smaller. The numerical method at the pathological parameters is behaving so badly that we actually see a negative $\hat{Y}$ value in the first integration's second sub-step.
	
	The following provides proofs for non-identifiability in general for linear ODEs when using the RK4 numerical method. In the case of the linear ODE we can leverage an explicit form of the RK4 algorithm in Equation \eqref{eqn_RK4} to get a polynomial in $\beta_0$ and $\beta_1$. We prove non-identifiability for solutions that satisfy:
	\[\frac{\beta_0}{\beta_1} = \alpha,
	\]
	which ensures that the asymptotic size is $\alpha$. Our numerical work suggests that in practice, the asymptotic size typically has a strong signal in data, and estimates of the parameters -- $\hat{\beta}_0$ and $\hat{\beta}_1$ -- tend to reflect this, with $\hat{\beta}_0/\hat{\beta}_1\approx \alpha$. By using the transformation $\beta_0 = \alpha\beta_1$ we derive a polynomial in $\beta_1$ for the numerical step in the following lemma. 
	
	\begin{lemma}\label{lem_LinearRK4}
		For a first order linear ODE of the form
		\[\frac{dY}{dt} = \beta_0 - \beta_1 Y(t),
		\]
		consider $\beta_0=\alpha \beta_1$ (that is, parameter combinations for which the asymptotic size is $\alpha$). A single step of the Runge-Kutta 4th order numerical integration is given by\begin{equation}\label{eqn_polynomialReduced}
			\hat{Y}(t_j + h) = Y(t_j) + (\alpha - Y(t_j))\bigg( h\beta_1 - \frac{h^2}{2}\beta_1^2 + \frac{h^3}{6}\beta_1^3 - \frac{h^4}{24}\beta_1^4\bigg)
		\end{equation}
		where $h$ is the step size and $\alpha = \beta_0/\beta_1$ is the asymptotic $Y$ value.
	\end{lemma}
	\begin{proof}
		The Runge-Kutta scheme for $Y(t)$ given $\frac{dY}{dt} = f(Y(t), t, \theta)$ is calculated as 
		\begin{equation}\label{supp_eqn_rk4}
			\hat{Y}(t_j + h) = Y(t_j) + \frac{h}{6}(k_1 + 2k_2 + 2k_3 + k4)
		\end{equation}
		where $h$ is the step size and $k$ values are estimates of the gradient at iterated points given by
		\begin{align*}
			k_1 =&\ f(Y(t_j), t_j, \theta),\\
			k_2 =&\ f(Y(t_j) + hk_1/2, t_j + h/2, \theta),\\
			k_3 =&\ f(Y(t_j) + hk_2/2, t_j + h/2, \theta),\\
			k_4 =&\ f(Y(t_j) + hk_3, t_j + h, \theta).\\
		\end{align*}
		
		First we calculate formulas for each $k_i$ that depend only on $\beta$s and $Y(t_j)$ by substituting in the function definition sequentially. We start with
		\[k_1 = \beta_0 - \beta_1 Y(t_j),
		\]
		which we substitute into $k_2$ to get 
		\begin{align*}
			k_2 =&\ \beta_0 - \beta_1\bigg(Y(t_j) + \frac{h}{2}k_1\big)\\
			=&\ \beta_0 - \beta_1\bigg(Y(t_j) + \frac{h}{2}(\beta_0 - \beta_1 Y(t_j))\bigg)\\
			=&\ \beta_0 -\beta_1\bigg(Y(t_j) + \frac{h}{2}\beta_0\bigg) + \beta_1^2\frac{h}{2}Y(t_j).
		\end{align*}
		Subsequent $k$s are 
		\begin{align*}
			k_3 =&\ \beta_0 -\beta_1 \bigg(Y(t_j) + \frac{h}{2}\beta_0\bigg)  +\beta_1^2\bigg(\frac{h}{2}Y(t_j) + \frac{h^2}{4}\beta_0\bigg) - \beta_1^3\frac{h^2}{4}Y(t_j),\\
			k_4 =&\ \beta_0 -\beta_1 (Y(t_j) + h\beta_0)  +\beta_1^2\bigg(h Y(t_j) + \frac{h^2}{2}\beta_0\bigg)\\
			&\ - \beta_1^3\bigg(\frac{h^2}{2}Y(t_j) + \frac{h^3}{4}\beta_0\bigg) + \beta_1^4\frac{h^4}{4}Y(t_j).
		\end{align*}
		To extract the polynomial we substitute the $k$s into Equation \ref{supp_eqn_rk4} and do some re-organising of terms to get
		\begin{align*}
			\hat{Y}(t_j + h) =&\ Y(t_j)  + \frac{h}{6}\Bigg(6\beta_0 - \beta_1\big(6Y(t_j) +3h\beta_0\big) + \beta_1^2\big(3hY(t_j) + h^2\beta_0\big)\nonumber \\
			&\qquad\qquad\qquad\quad - \beta_1^3\bigg(h^2 Y(t_j) + \frac{h^3}{4} \beta_0 \bigg) + \beta_1^4 \frac{h^3}{4}Y(t_j)\Bigg)\nonumber\\
			=&\ Y(t_j) + h\beta_0 - \beta_1\bigg(hY(t_j) + \frac{h^2}{2}\beta_0\bigg)  + \beta_1^2\bigg(\frac{h^2}{2}Y(t_j) + \frac{h^3}{6}\beta_0\bigg)\nonumber\\
			&\qquad\qquad\qquad\quad - \beta_1^3\bigg(\frac{h^3}{6}Y(t_j) + \frac{h^4}{24}\beta_0\bigg) +\beta_1^4 \frac{h^4}{24}Y(t_j).
		\end{align*}
		
		The linear DE with negative slope always converges to the asymptotic size $\beta_0/\beta_1 = \alpha$, and in fact if we make the substitution $\alpha\beta_1 = \beta_0$ we can reduce to a function of a single variable $\beta_1$ to get
		\[\hat{Y}(t_j + h) = Y(t_j) + (\alpha - Y(t_j))\bigg( h\beta_1 - \frac{h^2}{2}\beta_1^2 + \frac{h^3}{6}\beta_1^3 - \frac{h^4}{24}\beta_1^4\bigg)
		\]
		as required.
	\end{proof}
	
	In the longitudinal model, we are estimating $\beta_1$ values that minimise the difference between sizes over time based on the numerical projection using estimated parameters and the data arising from the analytic solution. In a zero error situation, this involves finding $\hat{\beta}_1$ values which produce numerical steps that match the analytic solution. That is, minimising the difference between $Y(t_j+h)-Y(t_j)$ and $\hat{Y}(t_j + h) - Y(t_j)$, where the former uses the true parameter $\beta_1$ and the analytic solution to reach $Y(t_j + h)$ while the latter uses the estimated parameter $\hat{\beta}_1$ and the numerical approximation. We need to distinguish the true parameter $\beta_1$ as used in the analytic solution from what we are estimating -- $\hat{\beta}_1$ -- which only forms part of the numerical step, hence the hat. We are assuming that both start at the same true $Y(t_j)$. Lemma \ref{lem_LinearRK4} gives us an algebraic form for one step, but we want a general form for the full longitudinal model with $n$ steps.
	
	\begin{proposition}\label{prop_NumSols}
		For a linear first order ODE of the form
		\[\frac{dY}{dt} = \beta_0 - \beta_1 Y(t)
		\]
		within a longitudinal model using $n$ steps of the Runge-Kutta 4th order algorithm, constrain estimates to fall along the line $\hat{\beta}_0=\alpha \hat{\beta}_1$. Then $\hat{\beta}_1$ solves
		\begin{equation}\label{eqn_beta1Poly}
			0 =  h\hat{\beta}_1 - \frac{h^2}{2}\hat{\beta}_1^2 + \frac{h^3}{6}\hat{\beta}_1^3 - \frac{h^4}{24}\hat{\beta}_1^4 + e^{-\beta_1 h} - 1.
		\end{equation}
	\end{proposition}
	\begin{proof}
		We start by looking at what happens for a single step, so set $n=1$. From Lemma \ref{lem_LinearRK4} we have
		\begin{equation*}
			\hat{Y}(t_j + h) = Y(t_j) + (\alpha - Y(t_j))f(\hat{\beta}_1, h)
		\end{equation*}
		where
		\[f(\hat{\beta}_1, h) =  h\hat{\beta}_1 - \frac{h^2}{2}\hat{\beta}_1^2 + \frac{h^3}{6}\hat{\beta}_1^3 - \frac{h^4}{24}\hat{\beta}_1^4
		\]
		for convenience, as it will not change throughout this proof. Values of $\hat{\beta}_1$ estimated on data will be those that match the numerical solution to the analytic one, giving
		\begin{equation}\label{eqn_PropPf1}
			Y(t_j + h) - Y(t_j) = (\alpha - Y(t_j))f(\hat{\beta}_1, h)
		\end{equation}
		where the left hand size is the true growth increment between $t_j$ and $t_j+h$ arising from the analytic solution which uses the true parameter $\beta_1$, and the right is the numerical estimate of the increment that depends on the estimate $\hat{\beta}_1$. 
		
		From the analytic solution in Equation \eqref{eqn_linearDESoln} we know that
		\begin{equation*}
			Y(t_j + h) = \alpha + (Y(t_j) - \alpha)e^{-\beta_1 (t_j + h - t_j)} = \alpha + (Y(t_j) - \alpha)e^{-\beta_1h},
		\end{equation*}
		which we can substitute into Equation \eqref{eqn_PropPf1} to get
		\begin{align}
			(\alpha - Y(t_j))f(\hat{\beta}_1, h) =&\ \alpha + (Y(t_j) - \alpha)e^{-\beta_1h} - Y(t_j)\nonumber\\
			=&\ (Y(t_j) - \alpha)e^{-\beta_1h} - (Y(t_j) - \alpha)\nonumber\\
			=&\ (Y(t_j) - \alpha)(e^{-\beta_1h}-1).\label{eqn_PropPf2}
		\end{align}
		When solving for $\hat{\beta}_1$ we re-arrange Equation \eqref{eqn_PropPf2} to get
		\begin{align}
			0 =&\ (\alpha - Y(t_j))f(\hat{\beta}_1, h) - (Y(t_j) - \alpha)(e^{-\beta_1h}-1)\nonumber\\
			=&\ (\alpha - Y(t_j))\big(f(\hat{\beta}_1, h) + e^{-\beta_1h}-1\big).\label{eqn_PropPf3}
		\end{align}
		As we assume $\alpha$ -- the asymptotic size -- is a constant, the only variable in Equation \eqref{eqn_PropPf3} is $\hat{\beta}_1$, so the solutions will be the same as those that solve
		\begin{equation}\label{eqn_PropPf10}
			f(\hat{\beta}_1, h) + e^{-\beta_1h}-1 = 0.
		\end{equation}
		
		The $n$-step case comes into play when we have a sequence of $Y(t_j)$ that we are fitting to, or are dividing up the time between two and applying RK4 repeatedly. Repeated applications of the algorithm is one way of seeking greater numerical accuracy through a smaller step size. Taking from Equation \eqref{lem_LinearRK4}, but extending to $n$ steps gives
		\begin{align*}
			\hat{Y}(t_j + nh) = Y(t_j) &+ \underbrace{(\alpha - Y(t_j))f(\hat{\beta}_1, h)}_{\text{step 1}}\\
			&+  \underbrace{(\alpha - Y(t_j+h))f(\hat{\beta}_1, h)}_{\text{step 2}}\\
			& + \ldots + \underbrace{(\alpha - Y(t_j+(n-1)h))f(\hat{\beta}_1, h)}_{\text{step n}}\\
			= Y(t_j) &+ f(\hat{\beta}_1, h)\Bigg(\sum_{i=0}^{n-1}\alpha - Y(t_j +ih) \Bigg).
		\end{align*}
		
		As in the 1-step case, the $n$-step model will seek $\hat{\beta}_1$ values such that $\hat{Y}(t_j + nh) = Y(t_j + nh)$ to match the data, so we want
		\begin{equation}\label{eqn_PropPf4}
			Y(t_j + nh) - Y(t_j) = f(\hat{\beta}_1, h)\Bigg(\sum_{i=0}^{n-1}\alpha - Y(t_j +ih) \Bigg).
		\end{equation}
		As in the 1-step case we take the left hand side (LHS) of Equation \eqref{eqn_PropPf4} and substitute in the analytic solution $Y(t_j + nh) = \alpha + (Y(t_j) - \alpha)e^{-\beta_1nh}$ to get
		\begin{equation}\label{eqn_PropPf5}
			\alpha + (Y(t_j) - \alpha)e^{-\beta_1nh} - Y(t_j) = (Y(t_j) - \alpha)(e^{-\beta_1nh}-1),
		\end{equation}
		but this time we need to factorise $e^{-\beta_1nh}-1$ into
		\begin{equation}\label{eqn_PropPf6}
			e^{-\beta_1nh}-1 = \Bigg(\sum_{i=0}^{n-1}e^{-\beta_1 ih} \Bigg)(e^{-\beta_1h} - 1),
		\end{equation}
		which is true because
		\[ \Bigg(\sum_{i=0}^{n-1}a^i \Bigg)(a - 1) = \sum_{i=1}^{n}a^i - \sum_{i=0}^{n-1}a^i = a^n - 1.
		\]
		Substituting Equation \eqref{eqn_PropPf6} into Equation \eqref{eqn_PropPf5} gives 
		\begin{equation}\label{eqn_PropPf7}
			\text{LHS} = (Y(t_j) - \alpha)\Bigg(\sum_{i=0}^{n-1}e^{-\beta_1 ih} \Bigg)(e^{-\beta_1h} - 1).
		\end{equation}
		
		Let us return to the right side of Equation \eqref{eqn_PropPf4}, and look at the argument within the sum. We substitute in the analytic solution to give
		\begin{align*}
			\alpha - Y(t_j + ih) =&\ \alpha - (\alpha + (Y(t_j) - \alpha)e^{-\beta_1 ih})\\
			=&\ - (Y(t_j) - \alpha)e^{-\beta_1 ih}\\
			=&\ (\alpha - Y(t_j))e^{-\beta_1 ih}
		\end{align*}
		which we substitute back in to give
		\begin{equation}\label{eqn_PropPf8}
			\text{RHS} = f(\hat{\beta}_1, h)(\alpha - Y(t_j))\Bigg(\sum_{i=0}^{n-1} e^{-\beta_1 ih} \Bigg).
		\end{equation}
		
		Recombining Equations \eqref{eqn_PropPf7} and \eqref{eqn_PropPf8}, we subtract Equation \eqref{eqn_PropPf7} from both sides to give
		\begin{align}
			0 =&\  f(\hat{\beta}_1, h)(\alpha - Y(t_j))\Bigg(\sum_{i=0}^{n-1} e^{-\beta_1 ih}\Bigg) - (Y(t_j) - \alpha)\Bigg(\sum_{i=0}^{n-1}e^{-\beta_1 ih} \Bigg)(e^{-\beta_1h} - 1)\nonumber\\
			=&\ (\alpha - Y(t_j))\Bigg(\sum_{i=0}^{n-1} e^{-\beta_1 ih}\Bigg)(f(\hat{\beta}_1, h) + e^{-\beta_1h} - 1). \label{eqn_PropPf9}
		\end{align}
		Equation \eqref{eqn_PropPf9} is 0 when
		\begin{equation}\label{eqn_PropPf11}
			0 = f(\hat{\beta}_1, h) + e^{-\beta_1h} - 1 = h\hat{\beta}_1 - \frac{h^2}{2}\hat{\beta}_1^2 + \frac{h^3}{6}\hat{\beta}_1^3 - \frac{h^4}{24}\hat{\beta}_1^4 + e^{-\beta_1 h} - 1,
		\end{equation}
		which is the same as Equation \eqref{eqn_PropPf10} from the 1-step case. As such, $\hat{\beta}_1$ solutions will be the same for all $n$. Equation \eqref{eqn_PropPf11} does not depend on $t_j$, so as long as $Y(t_j)\neq \alpha$ the solutions are independent of $Y(t_j)$ as well.
	\end{proof}
	
	Now that we have a general form for where $\hat{\beta}_1$ estimates come from, we can prove non-uniqueness from the polynomial that $\hat{\beta}_1$ solves.
	
	\begin{theorem}\label{thm_QuarticRts}
		For a linear first order ODE of the form
		\[\frac{dY}{dt} = \beta_0 - \beta_1 Y(t)
		\]
		where $\beta_0=\alpha \beta_1$,
		within a longitudinal model using $n$ steps of the Runge-Kutta 4th order algorithm, in general there are multiple $\hat{\beta}_1$ values that match the analytic solution for sizes over time.
	\end{theorem}
	\begin{proof}
		From Proposition \ref{prop_NumSols} we know that $\hat{\beta}_1$ values that produce a numerical projection that matches the analytic solution will solve
		\begin{equation*}
			0 = h\hat{\beta}_1 - \frac{h^2}{2}\hat{\beta}_1^2 + \frac{h^3}{6}\hat{\beta}_1^3 - \frac{h^4}{24}\hat{\beta}_1^4 + e^{-\beta_1 h} - 1
		\end{equation*}
		which is a polynomial in $\hat{\beta}_1$ of degree 4. The solutions can be explicitly calculated using the quartic root formula. 
		
		For some parameter values Equation \eqref{eqn_beta1Poly} may have a combination of real and complex roots, but as there must be at least one solution which approximates with the true (real) parameter $\beta_1$, we can exclude the case of all imaginary roots. If the discriminant of the polynomial in $\hat{\beta}_1$ is 0 there are some special cases where one real root (of some multiplicity) exists, that can be determined with further calculations when $h$ and $\beta_1$ are known. For example, for a given $h$ there will be one $\beta_1$ such that there is a single real root of multiplicity four given by
		\[\hat{\beta}_1 = -\frac{h^3}{6} \bigg(\frac{4h^4}{24}\bigg)^{-1} = -\frac{h^3}{6} \frac{24}{4h^4} = -\frac{1}{h},
		\]
		and $\hat{\beta}_1$ can be determined through calculation using the discriminant when $h$ is chosen \citep{Rees01021922}.
	\end{proof}
	
	One immediate consequence of these results is inaccuracy in the $\hat{\beta}_1$ value closest to the true $\beta_1$, due to the numerical approximation.
	\begin{corollary}\label{coro_Bias}
		For a linear first order ODE of the form
		\[\frac{dY}{dt} = \beta_0 - \beta_1 Y(t),\quad \beta_1\neq 0
		\]
		given true $\beta_0$ and $\beta_1$, within a longitudinal model using $n$ steps of the Runge-Kutta 4th order algorithm, the $\hat{\beta}_1$ closest to the true solution is not equal to it, that is, $\hat{\beta}_1\neq \beta_1$.
	\end{corollary}
	\begin{proof}
		We take 
		\begin{equation*}
			h\hat{\beta}_1 - \frac{h^2}{2}\hat{\beta}_1^2 + \frac{h^3}{6}\hat{\beta}_1^3 - \frac{h^4}{24}\hat{\beta}_1^4 + e^{-\beta_1 h} - 1
		\end{equation*}
		from Proposition \ref{prop_NumSols}. Set $\hat{\beta}_1 = \beta_1$ to give
		\[h\beta_1 - \frac{h^2}{2}\beta_1^2 + \frac{h^3}{6}\beta_1^3 - \frac{h^4}{24}\beta_1^4 + e^{-\beta_1 h} - 1,
		\]
		then we can exploit the series expansion of the exponential function
		\[e^x = \sum_{n=0}^\infty \frac{x^n}{n!}
		\]
		by setting $x = -h\beta_1$ to get
		\begin{align*}
			e^{- h\beta_1} - \bigg(1 - h\beta_1 + \frac{h^2\beta_1^2}{2} - \frac{h^3\beta_1^3}{6} + \frac{h^4\beta_1^4}{24} \bigg) =&\ \sum_{n=0}^\infty \frac{(-h\beta_1)^n}{n!} - \sum_{n=0}^4 \frac{(-h\beta_1)^n}{n!}\\
			=&\ \sum_{n=5}^\infty \frac{(-h\beta_1)^n}{n!}\\
			\neq&\ 0
		\end{align*}
		unless $h$ or $\beta_1$ is 0.
		
		As the true value $\hat{\beta_1} = \beta_1$ does not perfectly match the numeric solution to the analytic one, the estimate closest to $\beta_1$ that does solve Equation \eqref{eqn_beta1Poly} will not be $\hat{\beta_1} = \beta_1$.
	\end{proof}
	
	We suspect that Corollary \ref{coro_Bias} implies bias -- even for the best of the available estimates of $\beta_1$.
	
\begin{figure}[bth]
		\centering
		\includegraphics[width=0.6\linewidth]{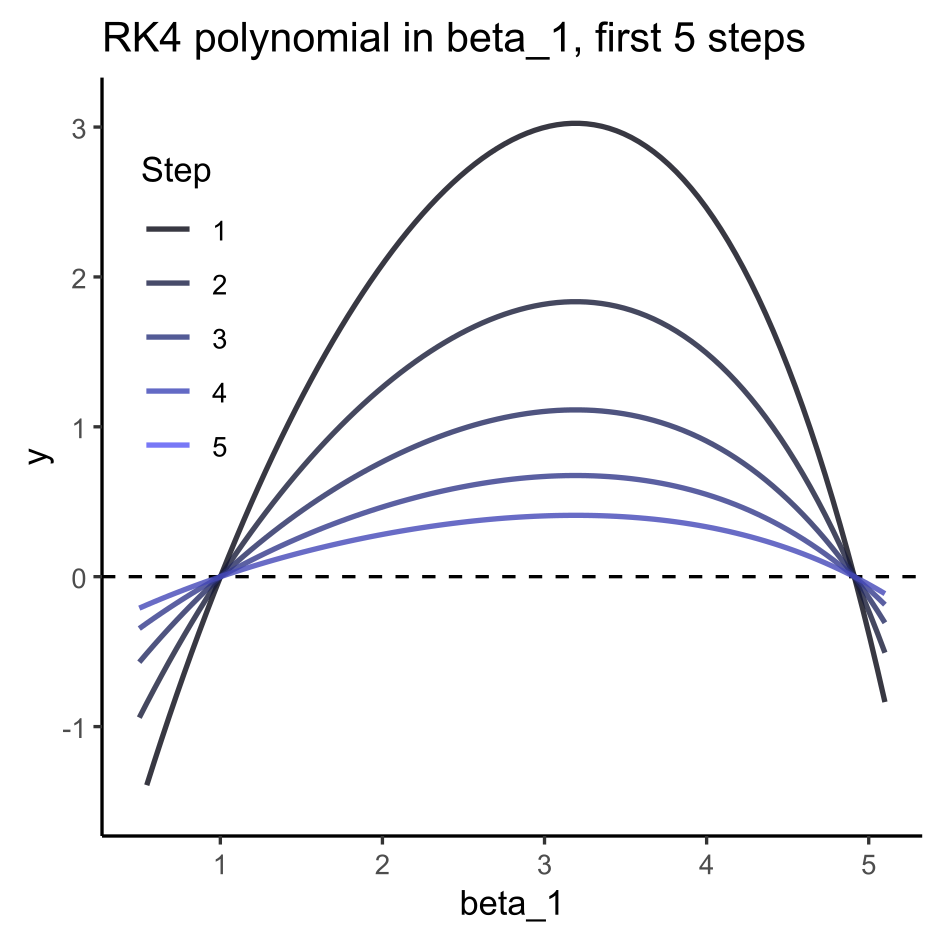}
		\caption{Plot of Equation \eqref{eqn_PropPf3}, where the $y$-axis replaces 0 on the left hand side, with parameters $h=1/2$, $\beta_1 = 1$, and $\alpha = 10$ for the first five steps. The intersections with 0 are at $\hat{\beta}_1 = 1.0008,\  4.9112$.}
		\label{fig_PolyZeros}
	\end{figure}

	We have described an example case where $\beta_1 = 1$, and $\beta_0 = 10$. The two real number solutions to Equation \eqref{eqn_beta1Poly} (the complex solutions do not appear because parameter estimation is restricted to real values) for those values are $\hat{\beta}_1 = 1.0008$ and $4.9112$, which is shown in Figure \ref{fig_PolyZeros} and aligns with the behaviour in Figure \ref{fig_Numerics}. The small difference from $\beta_1=1$ in the better estimate is to be expected given Corollary \ref{coro_Bias}.
	
	To conclude the theory section, the posterior distribution for $\beta_1$ and $\beta_0$ \emph{conditioned on the RK4 step} will have modes at the parameter values that produce $\hat{Y}(t_j)$ that match the analytic solution. The solutions to Equation \eqref{eqn_beta1Poly} gives posterior modes that cannot be distinguished by estimation based on a longitudinal model using the RK4 algorithm, which is precisely what non-identifiability means in this context. 
	
	Equation \ref{eqn_beta1Poly} is calculable directly because we have an analytic solution, explicit numerical method, and a straight-forward function. In other situations where either an analytic solution is unavailable, or if non-linearity makes an explicit step hostile to manipulation, simulation can provide empirical evidence of bimodality.
	
	\section{Empirical testing}\label{sec_Empirical}
	In conjunction with the theoretical results that prove non-identifiability, we tested how the problem manifests in practice when estimating parameters from simulated repeat observation data.
	
	\begin{figure}
		\centering
		\includegraphics[width=\linewidth]{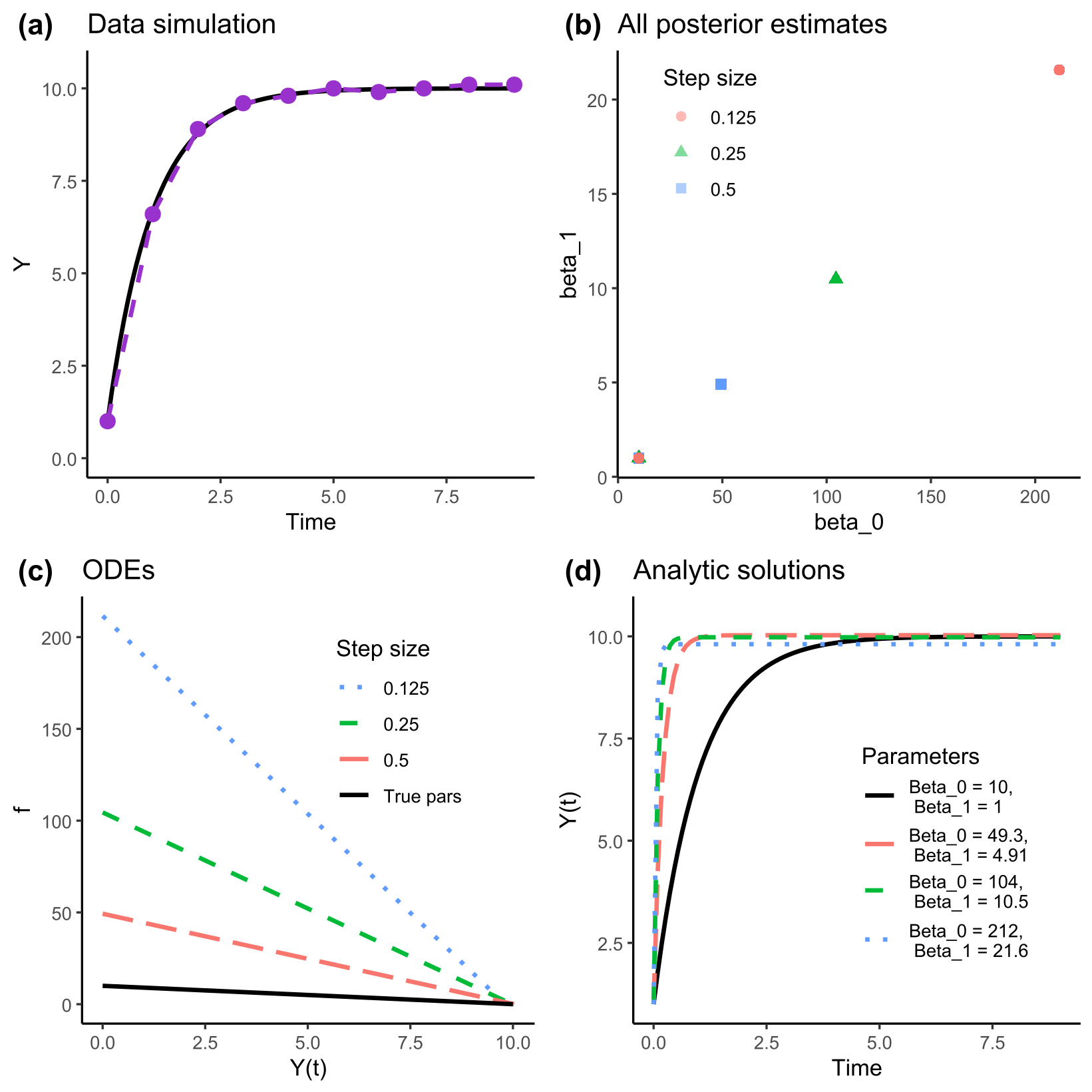}
		\caption{\textbf{(a)} gives the simulated data with measurement error and precision of 0.1 compared to the analytic solution for $\beta_0 = 10$, $\beta_1 = 1$. Outside of the specified situations with independent error for each chain, estimation is based on the same data. \textbf{(b)} is a scatter plot of all posterior estimates, good and bad, for the default priors across three step sizes using RK4. While there are a total of 30\,000 points on this plot, there is so little variation in them that they appear at just four points -- one near the true value, and for each step size, a second smaller cluster of points stacked around an erroneous value. \textbf{(c)} gives the erroneous differential equations and shows that all estimates converge to the same maximum size. \textbf{(d)} shows the analytic solutions using the erroneous parameter combinations to show that the erroneous values give sizes that converge to the asymptotic maximum much faster than the true solution.}
		\label{fig_DragonDE_Yt}
	\end{figure}
	
	\subsection{Aims}
	Our key question is: \emph{Do different integration methods have persistent bimodality?} We have demonstrated mathematical bimodality for RK4 with a step size of 0.5 so we test whether that shows up empirically as well. We also do empirical tests for RK4 with step sizes 0.25 and 0.125, RK45 with adaptive step size, and an analytic solution.
	
	We have further questions are addressed in the supplementary material:
	\begin{itemize}
		\item \emph{Can changes to estimation method limit the impact of bimodality?} We test different prior configurations for MCMC using the RK4 method with a step size of 0.5, and also test a deterministic optimisation algorithm L-BFGS using the default priors.
		\item \emph{Is there a model that provides accurate unimodal estimates?} We do simulation studies for the linear function and a highly non-linear model to test whether the chosen work-arounds -- an analytic implementation and RK45 with adaptive step size respectively -- have reliable results. 
	\end{itemize}
	
	\subsection{Methods}
	To address the questions we simulated data so we know the true values. We produce `true' data from the analytic solution Equation \eqref{eqn_linearDESoln} with parameters $\beta_0 = 10$, $\beta_1 = 1$ and initial size $Y(0) = 1$. The simulated survey structure is 10 size values $Y(t_j)$ with an interval of 1 unit of time between them, so $t_{j+1} - t_j = 1$. From the analytic sizes over time we get `observations' with error, rounded to precision 0.1, of the form
	\[y_j = \text{round}(Y(t_j) + \mathcal{N}(0, 0.1),\ 0.1).
	\]
	Figure \ref{fig_DragonDE_Yt}(a) shows sizes over time with error compared to the analytic solution. The simulation structure was chosen to behave like observations in centimetres with precision of \SI{1}{\milli\metre}, and measurement error with \SI{1}{\milli\metre} standard deviation. Results for the first two questions are based on running independent chains on the same set of 10 observations, so those results are conditioned on a single dataset. To test whether the implemented solution to bimodality is robust we simulate observations independently for each chain based on the same underlying $Y(t_j)$.
	
	Our three questions require a set of empirical tests. For each test we ran 10,000 independent chains \emph{on the same dataset}. In the case of MCMC this is 10,000 independent chains, while for the optimisation algorithm it is 10,000 runs each of which randomly selects a starting point in the parameter space. The large sample size was chosen because in initial testing we found that the second mode for the smallest RK4 step size was very rare.
	
	Estimation of $\beta_0$ and $\beta_1$ employs a single-individual version of the hierarchical Bayesian longitudinal model introduced in \cite{obrien2024allindividuals}, with implementation and workflow based on \texttt{hmde} (\url{https://github.com/traitecoevo/hmde}). As there is only one individual, the model lacks a `population' level, and instead we have fixed priors on the distributions for the two function parameters. Estimation is done through Markov Chain Monte Carlo sampling or a deterministic method (L-BFGS optimisation) implemented in Stan \citep{stan2022} via RStan \citep{rstan2019}. Code at \url{https://github.com/Tess-LaCoil/hmde-be-dragons} allows for complete reproduction of these results and further investigation as required.
	
	The fixed step sizes (0.5, 0.25, 0.125) were chosen for both providing an integer number of sub-steps to the observation period $t_{j+1} - t_j = 1$, and because larger step sizes had bad numerical behaviour in the proximity of the true parameter values.
	
	From each chain we extract the estimates $\hat{\beta}_0$ and $\hat{\beta}_1$ as the means of the posterior samples. We fit multivariate normal finite mixture models \citep{mclachlan2019finite} to the 10,000 estimates for each MCMC model using the \texttt{mixtools} package \citep{bengalia2009MixTools} which provides estimates of cluster means, variances, and mixture probabilities. To initialise the finite mixture algorithm, we classify clusters by taking points on either side of the sample mean for $\hat{\beta}_0$, giving
	\[g_1 = \{(\hat{\beta}_{0,j}, \hat{\beta}_{1,j}) : \hat{\beta}_{0,j} \leq \Bar{\hat{\beta}}_0\},\qquad g_2 = \{(\hat{\beta}_{0,j}, \hat{\beta}_{1,j}) : \hat{\beta}_{0,j} > \Bar{\hat{\beta}}_0\}.
	\]
	To avoid convergence problems we initialise the finite mixture algorithm with the sample means of the two initial clusters. The fitted mixture of normal distributions is then used to analyse the clustering behaviour, as we get cluster mean estimate vectors, covariance matrices, and cluster occurrence probability estimates. We also use scatter plots to look at the cluster marginal posterior parameter distributions. In the unimodal case we look at the mean of the estimated posterior parameters to check for evidence of large errors, and a 95\% posterior credible interval \citep{gelman1995bayesian} built from taking the central 95\% quantile interval of the posterior estimates to get an idea of spread. 
	
	To compare the effect of different priors we use the posterior mixtures models and estimates of the posterior density at the posterior modes. We calculate the prior density at each of the posterior modes to see how concentrated the prior is at each point, and for comparison to the posterior probability. If the prior has a large impact we would expect to see posterior probabilities that align with the relative size of the prior density. If the prior is not informative, differences in the prior density will not change the posterior probabilities.
	
	With the estimates in hand we can independently verify the numerical error using the \texttt{ode} function from the \texttt{deSolve} package \citep{soetaert2010deSolve}, which has its own implementation of the Runge-Kutta 4th order method (\texttt{rk4}). We plot the $\hat{Y}(t_j)$ numerical values for the true and estimated parameters with the step size used in the MCMC sampling and calculate the root mean squared error to show that the empirical results reflect the theoretical match between the forward projections. We use plots of the analytic solutions to show that the erroneous parameter combinations actually produce very different behaviour in the absence of numerical error.
	
\begin{figure}
		\centering
		\includegraphics[width=0.65\linewidth]{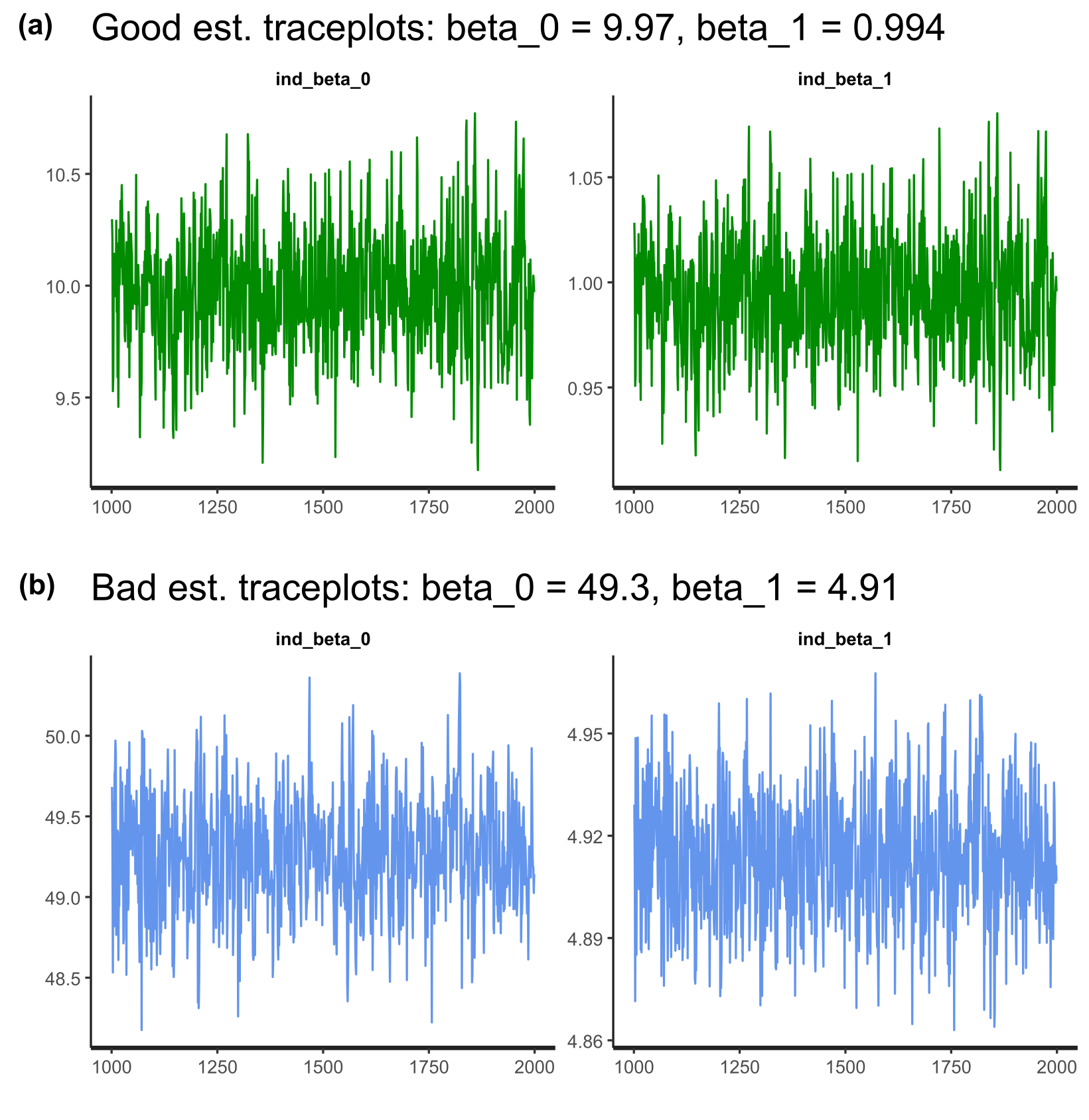}
		\caption{``Well-behaved'' chains converging to very different estimates of the parameters $\beta_0 = 10$, $\beta_1 = 1$. \textbf{(a)} produced good estimates while \textbf{(b)} converged to the erroneous parameter combination.}
		\label{fig_Traceplot}
	\end{figure}

	As we are running tens of thousands of simulations and looking for bad behaviour in the estimates we do not check diagnostics for every fit. In practice, consistent convergence of chains to different estimates is an indicator of posterior multi-modality and the numerical error may be a source of it. In Figure \ref{fig_Traceplot} we give examples of two chains fit with RK4 and a step size of 0.5 that appear to be well-behaved despite one of the two converging to the incorrect parameters, indicating that typical diagnostic plots alone may not be enough. 
	
	\section{Results}
	We see persistent bimodality across numerical methods with fixed step size. The location of the second mode depends on the numerical method. Changing the location of the parameter priors does not prevent convergence to the erroneous mode, while constraining the prior spread only eliminates the second mode when the standard deviation is too small to be of practical use. The analytic solution produces accurate, unimodal posteriors.
	
	\subsection{Numerical methods have persistent bimodality}
	All the step sizes tested with the RK4 algorithm had a second cluster of erroneous parameter estimates. As the step size shrinks, the erroneous cluster moves further away from the true values, and the probability of erroneous convergence decreases as shown in Table \ref{tab_Dragon_ErrorEst}. Figure \ref{fig_DragonDE_Yt} shows the cluster distance in Panel (b), where all 30,000 estimates are plotted together yet they appear as if single points due to the tiny within-cluster variance, as in Table \ref{tab_Dragon_ErrorEst}. There is a stack of points at (9.95, 0.991) which is the first cluster of estimates for all step sizes, and each step size has a second cluster spread further from that point as the step size decreases. The clusters all lie along a line preserving the asymptotic size at $\hat{\beta}_0/\hat{\beta}_1 \approx 10$, which is also visible in Figure \ref{fig_DragonDE_Yt}(c) and (d). Panel (c) shows Equation \eqref{eqn_linearDE} with parameters fit to the mean of each erroneous cluster compared to the true values, and Panel (d) shows the analytic solution Equation \eqref{eqn_linearDESoln} with the same. The erroneous cluster has much faster analytic solution convergence to the asymptotic size, but as we saw in Figure \ref{fig_Numerics}, that leads to bad numerical behaviour. 
	
\begin{figure}
		\centering
		\includegraphics[width=0.78\linewidth]{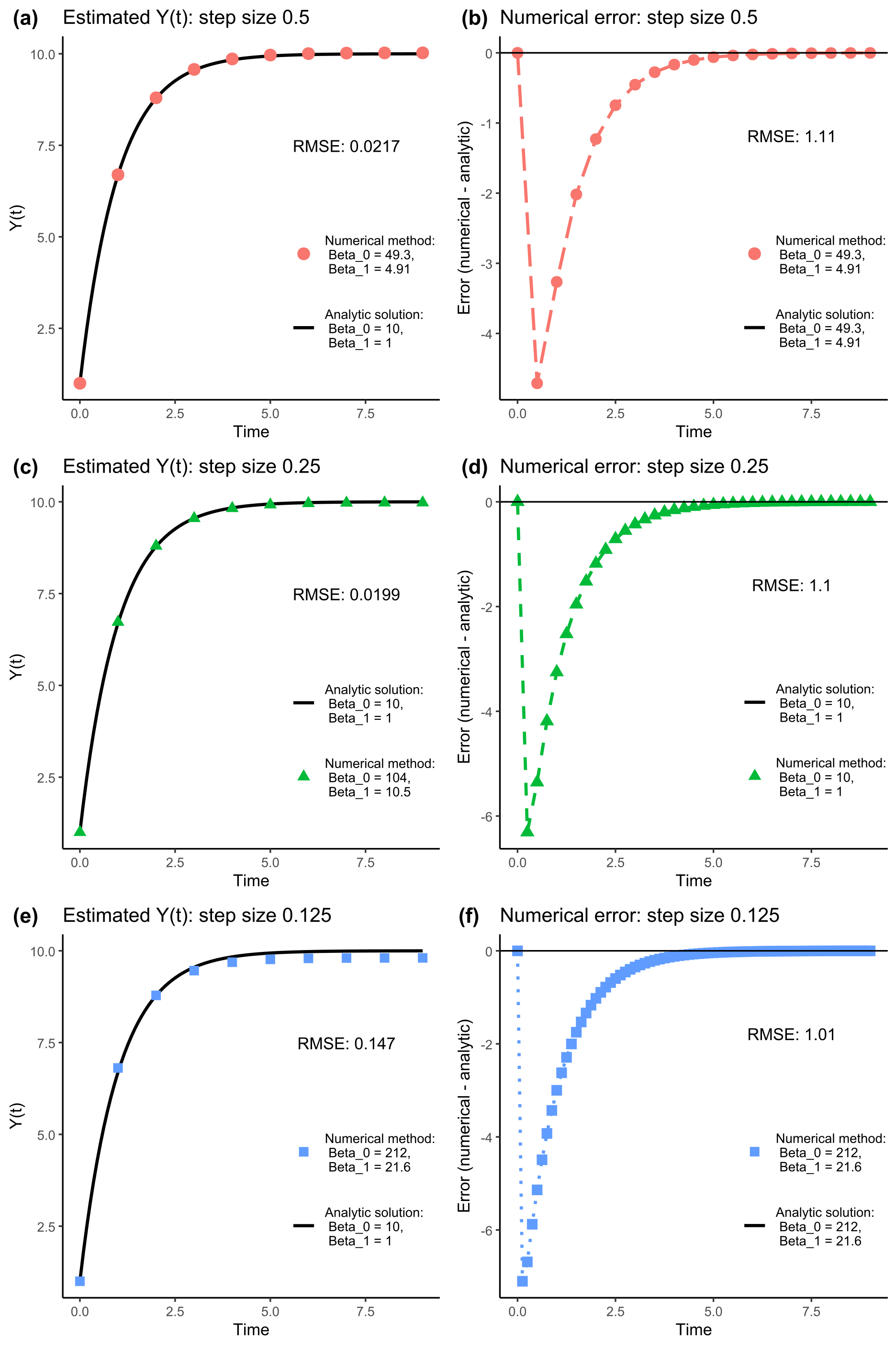}
		\caption{Estimated $\hat{Y}(t)$ from the linear model across three step sizes compared to the true linear model for simulated data, and numerical error. \textbf{(a)}, \textbf{(c)}, and \textbf{(e)} show that the estimated sizes over time for the erroneous cluster align closely with the analytic solution for the true $\beta$s across the step sizes. The slight difference in asymptotic size for step size 0.125 is likely due to the prior constraining $\hat{\beta}_0$. \textbf{(b)}, \textbf{(d)}, and \textbf{(f)} show the numerical error for each erroneous cluster, where the initial step has the largest error and then there is convergence to the same asymptotic size.}
		\label{fig_StepSizeYt}
	\end{figure}

	For each step size we give more detailed plots. Figure \ref{fig_StepSizeYt}(a), (c) and (e) show that the numerical estimates at the erroneous cluster strongly agree with the analytic solution for the true parameters, while (b), (d) and (f) shows how numerical error behaves badly by comparing the numerical estimates at the erroneous cluster to the analytic solution at the same parameter values.
	
\begin{figure}
		\centering
		\includegraphics[width=\linewidth]{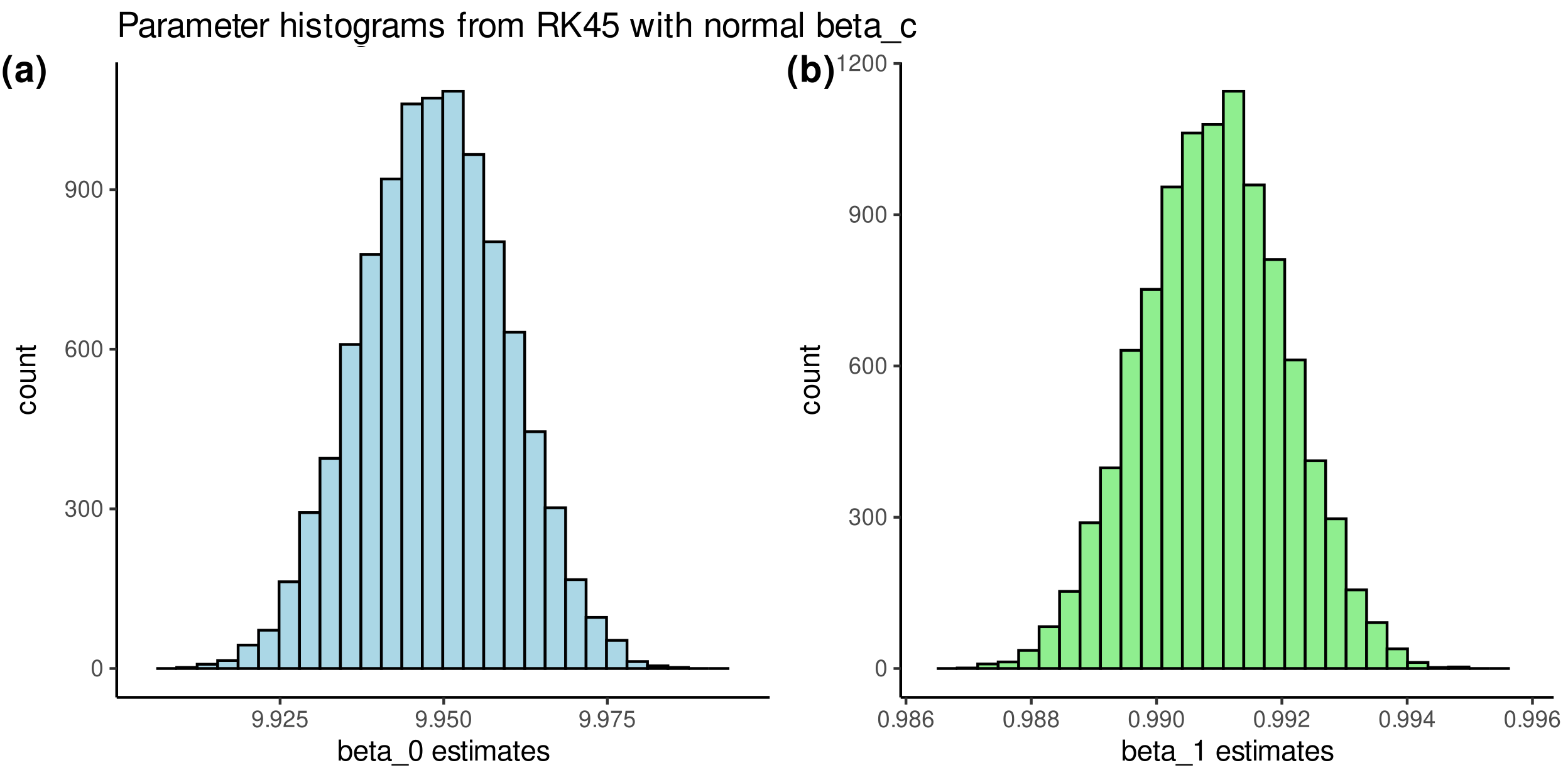}
		\caption{\textbf{(a)} and \textbf{(b)} show that the RK45 algorithm gives a unimodal posterior for both parameters when fit with a normal distribution prior on $\beta_c$.}
		\label{fig_RK45Analytic}
	\end{figure}

\begin{figure}
		\centering
		\includegraphics[width=\linewidth]{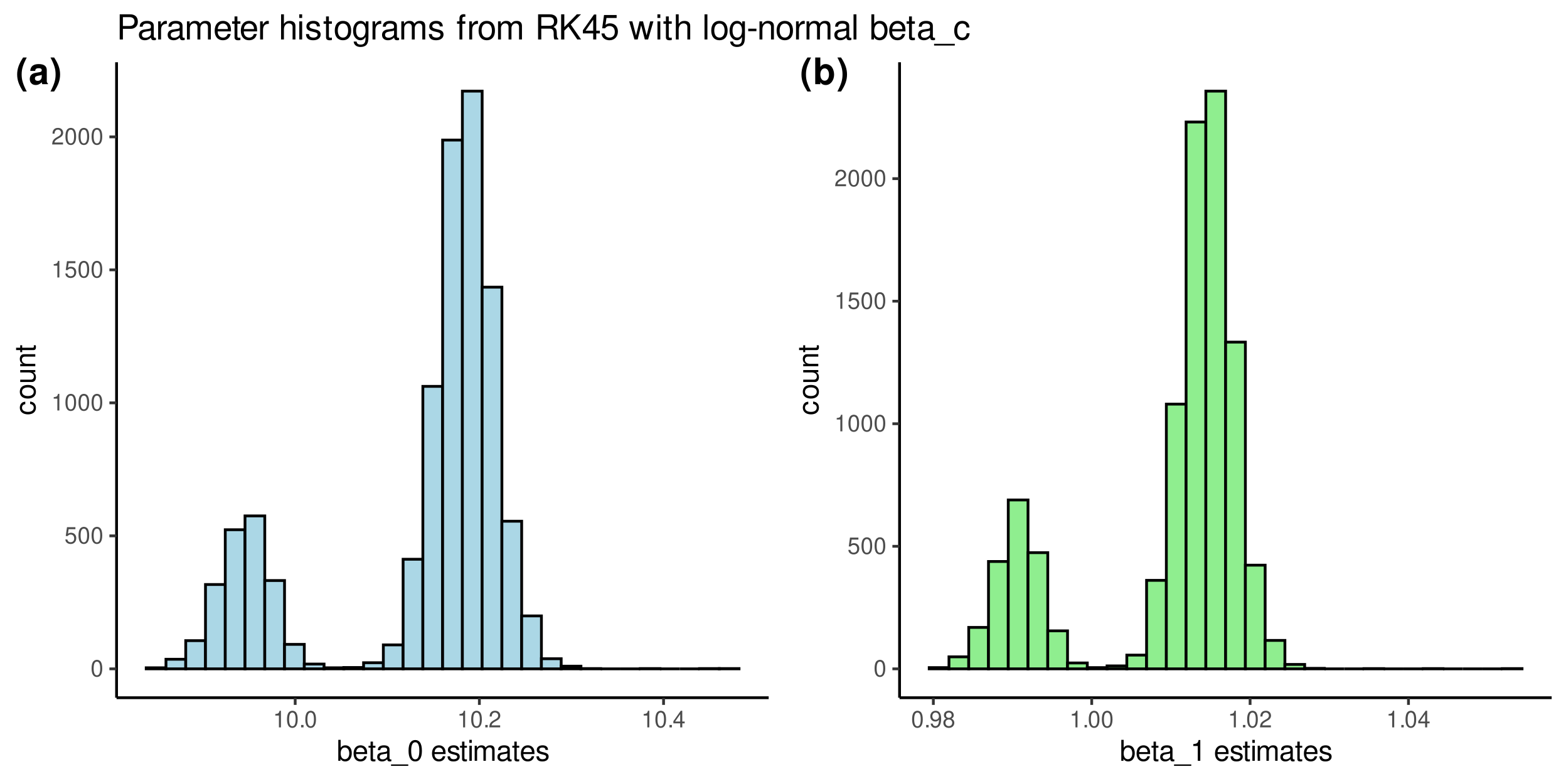}
		\caption{\textbf{(a)} and \textbf{(b)} show that bimodality can arise with the RK45 algorithm when $\beta_c$ has a log-normal prior, and in fact the second mode is more likely than the first.}
		\label{fig_RK45Bimodal}
	\end{figure}

	We tested a better numerical method as well -- RK45 with adaptive step size -- and found that under some prior specifications bimodality still occurred. Figure \ref{fig_RK45Analytic} shows unimodal posteriors for both the analytic and RK45 posteriors when $\beta_c$ has a normal prior, while Figure \ref{fig_RK45Bimodal} shows the bimodal posterior that arose when we was testing a log-normal distribution for $\beta_c$ with RK45, a problem that the analytic solution does not have. Log-normal priors are common so we include this example as a warning. Table \ref{tab_Dragon_ErrorEst} gives the mixture-model analysis for RK45. We do not consider the two clusters for $\beta_c\sim\log\mathcal{N}$ to be meaningfully different as they are so close together.
	
	\begin{sidewaystable}
		\renewcommand{\arraystretch}{1.2}
		\centering
		\caption{Posterior finite mixture model estimates for MCMC testing different numerical methods for the linear ODE. True parameter values are $\beta_0 = 10$ and $\beta_1 = 1$. The sample standard deviations of each cluster for each parameter are given in parentheses. Note that for each choice of step size, RK4 converged to one of two very different solutions.}
		\label{tab_Dragon_ErrorEst}
		\begin{tabular}{p{2.8cm}lcccccc}
			\toprule
			\multirow{2}{2.8cm}{\textbf{Numerical method}}&& \multicolumn{2}{c}{\textbf{Cluster 1}} & & \multicolumn{2}{c}{\textbf{Cluster 2}} & \textbf{Cluster 2}\\
			\cline{3-4}\cline{6-7}
			&\textbf{Step size} & $\mu_{\beta_0,1}\ (sd(\hat{\beta}_0,1))$ & $\mu_{\beta_1,1}\ (sd(\hat{\beta}_1,1))$ & & $\mu_{\beta_0,2}\ (sd(\hat{\beta}_0,2))$ & $\mu_{\beta_1,2}\ (sd(\hat{\beta}_1,2))$ & \textbf{prob.}\\
			\midrule
			\multirow{3}{2.8cm}{RK4 step size\\ analysis} & 0.5 & 9.956 (0.011) & 0.9917 (0.0012) && 49.26 (0.015) & 4.913 (0.00075) & 0.3195\\
			& 0.25 & 9.949 (0.011) & 0.991 (0.0012) && 104.5 (0.024) & 10.47 (0.00072) & 0.1044\\
			& 0.125 & 9.949 (0.011) & 0.9909 (0.0011) && 211.6 (0.038) & 21.57 (0.00067) & 0.0254\\
			\cline{2-8}
			RK45, $\beta_c\sim\mathcal{N}$ & Adaptive & 9.949 & 0.9909 && 9.949 & 0.9909  & 0.4729\\
			RK45, $\beta_c\sim\log\mathcal{N}$ & Adaptive & 9.945 & 0.9906 && 10.19 & 1.015  & 0.80\\
			\bottomrule
		\end{tabular}
	\end{sidewaystable}   
	
	\subsection{Implementing the analytic solution produces accurate, unimodal estimates}
	The simulation tests for the linear model with an analytic solution produced a unimodal distribution that was centred within 0.3\% of the true values as shown in Table \ref{tab_UnbiasedEst} which is a negligible empirical bias. Using the analytic solution means we do not introduce the non-identifiability that arises from numerical error so this result is in line with our theory.
	
	\begin{table}
		\caption{The linear model with an analytic solution shows minimal evidence of bias.}
		\label{tab_UnbiasedEst}
		\renewcommand{\arraystretch}{1.3}
		\centering
		\begin{tabular}{llccc}
			\toprule
			\textbf{Model} & \textbf{Parameter} & \textbf{True value} & \textbf{Mean est.} & \textbf{95\% CI}\\
			\midrule
			\multirow{2}{*}{Linear}&$\beta_0$ & 10  & 10.02 & $(9.484,\ 10.57)$\\
			& $\beta_1$ & 1 & 1.002 & $(0.9445,\ 1.062)$\\
			\bottomrule
		\end{tabular}
	\end{table}
	
	\section{Discussion}
	This paper provides theoretical and empirical evidence of a multi-modal posterior arising from numerical error in a longitudinal model for a linear first order ODE. One posterior mode aligns with good estimates of the true parameter values for simulated data, and at that mode the numerical method is reasonably well behaved. The other mode has extremely bad estimates of the true parameters, and badly behaved numerics as well. The existence of non-identifiability for RK4 in the case of a linear ODE demonstrates that care must be taken when using numerical methods for longitudinal models. 
	
	We show that the linear model can use an analytic solution to avoid the numerical pathology. The simulation testing procedure is more generally applicable and should form part of longitudinal model implementation if there is evidence of bad posterior behaviour such as divergent chains. It is also possible to test estimates from real world data after the fact by using a more accurate numerical method than the one implemented in the model.
	
	Using numerical theory, we proved that in general, the RK4 algorithm in a longitudinal model for linear first-order ODEs with negative slope will produce multi-modal results, and demonstrate with a specific ODE. In the empirical results, we see two clusters of parameter estimates that produce statistically indistinguishable $\hat{Y}(t_j)$. The empirical parameter estimates align closely with the theoretical results.
	
	For the end user, the process of testing common ways to adapt a model and prevent problems will hopefully be demonstrative. We tested different step sizes for the numerical method, different priors, estimation processes, and implementations of the longitudinal structure. Identifying that the divergent chains problem originated from the numerical method was difficult, but the symptom was distinct once recognised: independent chains converging to different, but consistent, parameter values, one of which was a good estimate of the true parameters. If this behaviour is observed in a longitudinal model using numerical methods, check how the numerics behave at the different parameter values. Test both the numerical method as implemented in the longitudinal model, and a (computationally more expensive) more accurate method as well. 
	
	We demonstrated the bimodal posterior for Equation \eqref{eqn_linearDE}, and show that the behaviour can occur for different step sizes and numerical methods. Where the second mode is located depends on the numerical method and step size, as which combination of bad parameters gives `correct' $\hat{Y}(t_j)$ depends on both those factors. Changes to the prior distribution were unable to appropriately constrain the second mode, but implementing the analytic solution instead of using a numerical method eliminated the erroneous parameters in MCMC sampling. 
	
	The persistence of bimodal posteriors due to numerical error indicate that choice of integration method matters beyond bias in estimating the posterior parameters. Bias has been tested in \cite{agapiou2014analysis} and \cite{obrien2024allindividuals}. We are able to select integration methods that produce unimodal estimates with negligible bias. For the linear model we used the analytic solution.
	
	Our testing of different priors is distinct from the method recommended by \cite{gabry2019visualization}. We seek to have default priors that resemble common choices to show that bimodality occurs, then test a specific set of priors to see if the problem can be averted. The choice of $\beta_1 \sim \log\mathcal{N}(0,2)$ and $\beta_c\sim\mathcal{N}(1,2)$ would likely be `vague' under the taxonomy proposed by \cite{banner2020use}, though the standard deviation of 2 is much smaller than some choices for flat priors. Instead, the default standard deviation represents what could be plausible given the magnitude of measurement (on the order of 1 to \SI{10}{\centi\metre}) and the observed growth rates with a large-ish variance. The default priors are less important as we are not doing inference on data, W want to see if prior selection can constrain bimodality. We found that location did not, and the prior standard deviation needed to be unreasonably small to eliminate the second mode, far smaller than would ever be used in practice. We note that changing the prior distribution for the RK45 method from $\beta_c\sim\log\mathcal{N}(0,2)$ to $\mathcal{N}(1,2)$ did eliminate the second mode, so there may be specific cases where prior distribution choice works. The failure to eliminate the second mode with prior specification for the RK4 method is because the priors do not address the actual source of the problem. 
	
	Estimation methods are a consideration for models for both material (computation time) and theoretical reasons.MCMC sampling requires the numerical calculations to be run thousands of times so while an erroneous cluster can be identified after the fact with a high precision (and computationally expensive) numerical method, implementing that directly in the longitudinal model is not so feasible. That problems can be identified after the fact is helpful at least. In some cases multiple chains run in parallel can detect a bimodal posterior if they converge to different points, which is how we first observed the issue as tweaks to the von Bertalanffy model with RK4 failed to avoid the occasional chain converging to a secondary mode. Working with a small number of parallel chains is less reliable as a means of identifying posterior bimodality than large-scale simulation testing as we have done in this paper, particularly when the probability of converging to the second mode is small as for the linear model with step size 0.125. 
	
	If using empirical testing alone, MCMC may fail to detect numerical problems in parts of the parameter space, an issue pointed to by \cite{rannala2002identifiability} in a different context. If the user's goal is to get a good estimate and not be waylaid by misbehaving numerics, that is a good thing. If the aim is to determine where numerical methods fail to work for the specified model, a more sensitive method such as L-BFGS may be better for scoping out the posterior. There are alternate algorithms around, including in the inverse problem literature more broadly. \cite{bardsley2014randomize} uses a combination of random sampling and deterministic optimisation for example. Methods proposed by \cite{pompe2020framework, syed2022non} and the annealed leap-point sampler (ALPS) \citep{tawn2021annealed, roberts2022skew} are intended for multi-modal posteriors, but are not readily accessible to users. As the numerical integration is separate to the estimation structure, the suggestion in \cite{lele2009bayesian} that frequentist likelihood maximisation methods may be preferred is not a solution for these situations.
	
	The biggest consequence of this form of non-identifiability in Bayesian inverse methods that use numerical integration is that even if a reasonable set of `true' parameters is known, stability for the numerical methods matters for potentially a very large region of the parameter space. For the smallest step size we tried, the Euclidean distance between the good and bad modes was 217, and we found that constraining the priors did not meaningfully affect the posterior mode probability for the larger step size. 
	
	The numerical pathologies we found are adjacent to those addressed in the typical Bayesian workflow \citep{gelman2020bayesian}, as the error comes from the longitudinal structure that is not part of most models. The simulation testing that we propose for posterior bimodality is relatively easy to implement, compared to hard-core numerical analysis for example, and can be done prior to large-scale dataset estimation. 
	
	We have shown that simulation can detect bimodality, but even a unimodal posterior should be checked with a better numerical method after sampling. If the numerical method in the model behaves badly at the estimated parameters, the estimates cannot be trusted. In a more general setting, when using models that rely on numerical methods, numerical error should be considered alongside other sources such as measurement error \citep{cotter2010approximation}.
	
	Non-identifiability arising from numerical error in a longitudinal model does not appear to be addressed elsewhere in the identifiability literature. More attention is paid to situations where the data is insufficient to uniquely specify estimates from a theoretical standpoint, whether due to too large a parameter space for the observations \citep{gelfand1999identifiability} or large measurement uncertainty \citep{auger2016state}. \cite{gelman2020bayesian} Section 11 addresses multi-modality that arises from orbit periodicity where there are not enough observations to select a specific value. Our problem arises in the longitudinal part of the model structure, but the literature around model structure tends to focus on over-parameterisation \citep{rannala2002identifiability}, or prior specification \citep{gelfand1999identifiability}, neither of which are the problem here. As such, we hope this case study can expand the known issues with identifiability in inverse problems and Bayesian methods, and provide answers for someone else who has gotten stuck on a similar issue.
	
	\section{Conclusions}
	Here we document the discovery and investigation of a novel source of non-identifiability in Bayesian inverse methods. We have shown that posterior bimodality can arise from an interaction between numerical integration error in a longitudinal model and MCMC sampling of integrand parameters. 
	
	For a known troublemaker ODE -- Equation \eqref{eqn_linearDE} -- posterior bimodality was persistent across three step sizes for RK4. Only an analytic solution implemented in the longitudinal model was able to correct the issue.
	
	As Bayesian inverse methods develop, knowing where and how numerical error can introduce pathologies is essential to good estimation. We have demonstrated one such problem and solution, in a manner that is hopefully useful elsewhere.
	
	\newpage
	\section*{Acknowledgments}
		The first author would like to thank Alex Mundey for his help with working out the numerical scheme. The authors would like to thank the referees and editors for their helpful feedback.
	
	\section*{Funding}
		The first author was supported by a Research Training Program scholarship from the Australian Government -- Department of Education.

	\bibliographystyle{apalike}
	\bibliography{numerics}

\begin{thebibliography}{}

\bibitem[Agapiou et~al., 2014]{agapiou2014analysis}
Agapiou, S., Bardsley, J.~M., Papaspiliopoulos, O., and Stuart, A.~M. (2014).
\newblock Analysis of the {Gibbs} sampler for hierarchical inverse problems.
\newblock {\em SIAM/ASA Journal on Uncertainty Quantification}, 2(1):511--544.

\bibitem[Auger-M{\'e}th{\'e} et~al., 2016]{auger2016state}
Auger-M{\'e}th{\'e}, M., Field, C., Albertsen, C.~M., Derocher, A.~E., Lewis,
  M.~A., Jonsen, I.~D., and Mills~Flemming, J. (2016).
\newblock State-space models’ dirty little secrets: even simple linear
  {Gaussian} models can have estimation problems.
\newblock {\em Scientific Reports}, 6(1):26677.

\bibitem[Banner et~al., 2020]{banner2020use}
Banner, K.~M., Irvine, K.~M., and Rodhouse, T.~J. (2020).
\newblock The use of {Bayesian} priors in ecology: The good, the bad and the
  not great.
\newblock {\em Methods in Ecology and Evolution}, 11(8):882--889.

\bibitem[Bardsley et~al., 2014]{bardsley2014randomize}
Bardsley, J.~M., Solonen, A., Haario, H., and Laine, M. (2014).
\newblock Randomize-then-optimize: A method for sampling from posterior
  distributions in nonlinear inverse problems.
\newblock {\em SIAM Journal on Scientific Computing}, 36(4):A1895--A1910.

\bibitem[Benaglia et~al., 2009]{bengalia2009MixTools}
Benaglia, T., Chauveau, D., Hunter, D.~R., and Young, D. (2009).
\newblock {mixtools}: An {R} package for analyzing finite mixture models.
\newblock {\em Journal of Statistical Software}, 32(6):1--29.

\bibitem[Butcher, 2016]{butcher2016numerical}
Butcher, J.~C. (2016).
\newblock {\em Numerical methods for ordinary differential equations}.
\newblock John Wiley \& Sons, third edition.

\bibitem[Canham et~al., 2004]{canham2004neighborhood}
Canham, C.~D., LePage, P.~T., and Coates, K.~D. (2004).
\newblock A neighborhood analysis of canopy tree competition: effects of
  shading versus crowding.
\newblock {\em Canadian Journal of Forest Research}, 34(4):778--787.

\bibitem[Cockayne et~al., 2019]{cockayne2019bayesian}
Cockayne, J., Oates, C.~J., Sullivan, T.~J., and Girolami, M. (2019).
\newblock Bayesian probabilistic numerical methods.
\newblock {\em SIAM Review}, 61(4):756--789.

\bibitem[Condit, 1998]{condit1998tropical}
Condit, R. (1998).
\newblock {\em Tropical forest census plots: methods and results from Barro
  Colorado Island, Panama and a comparison with other plots}.
\newblock Springer Science \& Business Media.

\bibitem[Cotter et~al., 2010]{cotter2010approximation}
Cotter, S.~L., Dashti, M., and Stuart, A.~M. (2010).
\newblock Approximation of {Bayesian} inverse problems for {PDE}s.
\newblock {\em SIAM Journal on Numerical Analysis}, 48(1):322--345.

\bibitem[Falster et~al., 2017]{falster2017multitrait}
Falster, D.~S., Br{\"a}nnstr{\"o}m, {\AA}., Westoby, M., and Dieckmann, U.
  (2017).
\newblock Multitrait successional forest dynamics enable diverse competitive
  coexistence.
\newblock {\em Proceedings of the National Academy of Sciences},
  114(13):E2719--E2728.

\bibitem[Gabry et~al., 2019]{gabry2019visualization}
Gabry, J., Simpson, D., Vehtari, A., Betancourt, M., and Gelman, A. (2019).
\newblock Visualization in {Bayesian} workflow.
\newblock {\em Journal of the Royal Statistical Society Series A: Statistics in
  Society}, 182(2):389--402.

\bibitem[Gelfand and Sahu, 1999]{gelfand1999identifiability}
Gelfand, A.~E. and Sahu, S.~K. (1999).
\newblock Identifiability, improper priors, and {Gibbs} sampling for
  generalized linear models.
\newblock {\em Journal of the American Statistical Association},
  94(445):247--253.

\bibitem[Gelman et~al., 2021]{gelman1995bayesian}
Gelman, A., Carlin, J.~B., Stern, H.~S., and Rubin, D.~B. (2021).
\newblock {\em Bayesian Data Analysis}.
\newblock CRC Press, 3rd april 2021 edition.

\bibitem[Gelman et~al., 2020]{gelman2020bayesian}
Gelman, A., Vehtari, A., Simpson, D., Margossian, C.~C., Carpenter, B., Yao,
  Y., Kennedy, L., Gabry, J., B{\"u}rkner, P.-C., and Modr{\'a}k, M. (2020).
\newblock Bayesian workflow.
\newblock {\em arXiv preprint arXiv:2011.01808}.

\bibitem[Iida et~al., 2014]{iida2014linking}
Iida, Y., Kohyama, T.~S., Swenson, N.~G., Su, S.-H., Chen, C.-T., Chiang,
  J.-M., and Sun, I.-F. (2014).
\newblock Linking functional traits and demographic rates in a subtropical tree
  community: the importance of size dependency.
\newblock {\em Journal of Ecology}, 102(3):641--650.

\bibitem[{Karline Soetaert} et~al., 2010]{soetaert2010deSolve}
{Karline Soetaert}, {Thomas Petzoldt}, and {R. Woodrow Setzer} (2010).
\newblock Solving differential equations in {R}: Package de{S}olve.
\newblock {\em Journal of Statistical Software}, 33(9):1--25.

\bibitem[Latz, 2023]{latz2023bayesian}
Latz, J. (2023).
\newblock Bayesian inverse problems are usually well-posed.
\newblock {\em SIAM Review}, 65(3):831--865.

\bibitem[Lele and Dennis, 2009]{lele2009bayesian}
Lele, S.~R. and Dennis, B. (2009).
\newblock Bayesian methods for hierarchical models: are ecologists making a
  {Faustian} bargain?
\newblock {\em Ecological Applications}, 19(3):581--584.

\bibitem[McLachlan et~al., 2019]{mclachlan2019finite}
McLachlan, G.~J., Lee, S.~X., and Rathnayake, S.~I. (2019).
\newblock Finite mixture models.
\newblock {\em Annual Review of Statistics and its Application}, 6(1):355--378.

\bibitem[Modr{\'a}k et~al., 2023]{modrak2023simulation}
Modr{\'a}k, M., Moon, A.~H., Kim, S., B{\"u}rkner, P., Huurre, N.,
  Faltejskov{\'a}, K., Gelman, A., and Vehtari, A. (2023).
\newblock Simulation-based calibration checking for {Bayesian} computation: The
  choice of test quantities shapes sensitivity.
\newblock {\em Bayesian Analysis}, 1(1):1--28.

\bibitem[O'Brien et~al., 2024]{obrien2024allindividuals}
O'Brien, T., Warton, D., and Falster, D. (2024).
\newblock Yes, they're all individuals: Hierarchical models for repeat survey
  data improve estimates of tree growth and sie.
\newblock {\em Methods in Ecology and Evolution}, 16(1).

\bibitem[Pompe et~al., 2020]{pompe2020framework}
Pompe, E., Holmes, C., and {\L}atuszy{\'n}ski, K. (2020).
\newblock A framework for adaptive {MCMC} targeting multimodal distributions.
\newblock {\em The Annals of Statistics}, 48(5):2930--2952.

\bibitem[Rannala, 2002]{rannala2002identifiability}
Rannala, B. (2002).
\newblock Identifiability of parameters in {MCMC} {Bayesian} inference of
  phylogeny.
\newblock {\em Systematic Biology}, 51(5):754--760.

\bibitem[Rees, 1922]{Rees01021922}
Rees, E.~L. (1922).
\newblock Graphical discussion of the roots of a quartic equation.
\newblock {\em The American Mathematical Monthly}, 29(2):51--55.

\bibitem[Roberts et~al., 2022]{roberts2022skew}
Roberts, G.~O., Rosenthal, J.~S., and Tawn, N.~G. (2022).
\newblock Skew {Brownian} motion and complexity of the {ALPS} algorithm.
\newblock {\em Journal of Applied Probability}, 59(3):777--796.

\bibitem[Rothenberg, 1971]{rothenberg1971identification}
Rothenberg, T.~J. (1971).
\newblock Identification in parametric models.
\newblock {\em Econometrica: Journal of the Econometric Society}, pages
  577--591.

\bibitem[{Stan Development Team}, 2019]{rstan2019}
{Stan Development Team} (2019).
\newblock {RStan}: the {R} interface to {Stan}.
\newblock R package version 2.19.2.

\bibitem[{Stan Development Team}, 2022]{stan2022}
{Stan Development Team} (2022).
\newblock {Stan Modeling Language Users Guide and Reference Manual}.
\newblock Version 2.33.

\bibitem[Syed et~al., 2022]{syed2022non}
Syed, S., Bouchard-C{\^o}t{\'e}, A., Deligiannidis, G., and Doucet, A. (2022).
\newblock Non-reversible parallel tempering: a scalable highly parallel {MCMC}
  scheme.
\newblock {\em Journal of the Royal Statistical Society Series B: Statistical
  Methodology}, 84(2):321--350.

\bibitem[Talts et~al., 2020]{talts2020validating}
Talts, S., Betancourt, M., Simpson, D., Vehtari, A., and Gelman, A. (2020).
\newblock Validating {Bayesian} inference algorithms with simulation-based
  calibration.
\newblock arXiv preprint arXiv:1804.06788.

\bibitem[Tawn et~al., 2021]{tawn2021annealed}
Tawn, N.~G., Moores, M.~T., and Roberts, G.~O. (2021).
\newblock Annealed leap-point sampler for multimodal target distributions.
\newblock {\em arXiv preprint arXiv:2112.12908}.

\bibitem[Von~Bertalanffy, 1938]{von1938quantitative}
Von~Bertalanffy, L. (1938).
\newblock A quantitative theory of organic growth (inquiries on growth laws.
  ii).
\newblock {\em Human Biology}, 10(2):181--213.

\end{thebibliography}
	
\end{document}